\documentclass{aa}

\usepackage{graphicx}
\usepackage[varg]{txfonts}
\usepackage{natbib}
\usepackage{twoopt}
\usepackage[breaklinks=true]{hyperref}
\usepackage{xspace}

\hypersetup{colorlinks=true,linkcolor=blue,citecolor=blue,urlcolor=blue}

\bibpunct{(}{)}{;}{a}{}{,}             
\makeatletter
  \newcommandtwoopt{\citeads}[3][][]{\href{http://adsabs.harvard.edu/abs/#3}%
    {\def\hyper@linkstart##1##2{}%
     \let\hyper@linkend\@empty\citealp[#1][#2]{#3}}}
  \newcommandtwoopt{\citepads}[3][][]{\href{http://adsabs.harvard.edu/abs/#3}%
    {\def\hyper@linkstart##1##2{}%
     \let\hyper@linkend\@empty\citep[#1][#2]{#3}}}
  \newcommandtwoopt{\citetads}[3][][]{\href{http://adsabs.harvard.edu/abs/#3}%
    {\def\hyper@linkstart##1##2{}%
     \let\hyper@linkend\@empty\citet[#1][#2]{#3}}}
  \newcommandtwoopt{\citeyearads}[3][][]%
    {\href{http://adsabs.harvard.edu/abs/#3}
    {\def\hyper@linkstart##1##2{}%
     \let\hyper@linkend\@empty\citeyear[#1][#2]{#3}}}
\makeatother

\DeclareMathOperator{\dif}{d \!}

\providecommand\gaia{\textit{Gaia}\xspace}
\providecommand\gdr[1]{\textit{Gaia}~DR#1\xspace}

\providecommand{\kms}{{\,\mathrm{km\,s^{-1}}}}
\providecommand\figref[1]{Fig.~\ref{#1}}
\providecommand\secref[1]{Sect.~\ref{#1}}
\providecommand\equref[1]{Eq.~\eqref{#1}}
\providecommand\equrefalt[1]{Equation~\eqref{#1}}

\providecommand\vect[1]{\ensuremath{{\boldsymbol{#1}}}}
\providecommand\sysA{\ensuremath{{\cal A}}\xspace}
\providecommand\sysB{\ensuremath{{\cal B}}\xspace}
\providecommand\ua{\ensuremath{\vect{u}_{\sysA}}\xspace}
\providecommand\ub{\ensuremath{\vect{u}_{\sysB}}\xspace}
\providecommand\du{\ensuremath{\delta\vect{u}}\xspace}
\providecommand\vb{\ensuremath{\vect{V}}\xspace}
\providecommand\vssbc{\ensuremath{\vect{v}_{\rm c}}\xspace}
\providecommand\assbc{\ensuremath{\vect{a}_{\rm c}}\xspace}
\providecommand\vssbg{\ensuremath{\vect{v}_{\rm g}}\xspace}

\providecommand\inprod[2]{\ensuremath{{#1}\cdot{#2}}}
\providecommand\inprodp[2]{\ensuremath{\left({#1}\cdot{#2}\right)}}

\let\linenumbers\nolinenumbers\nolinenumbers

\begin{document}

\title{The \gaia Astrometric Catalogue and Secular Aberration Drift in Proper Motions}

\author{A.G.A. Brown\inst{1} \and U. Bastian\inst{2} \and S. Klioner\inst{3}}

\institute{Leiden Observatory, Leiden University, Einsteinweg 55, 2333 CC Leiden, The Netherlands\\
    \email{brown@strw.leidenuniv.nl}
    \and
    Astronomisches Rechen-Institut, Zentrum f\"{ u}r Astronomie der Universit\"{ a}t Heidelberg, M\"{ o}nchhofstr.\
    12--14, 69120 Heidelberg, Germany
    \and
    Lohrmann Observatory, Technische Universit\"{ a}t Dresden, Mommsenstra{\ss}e 13, 01062 Dresden, Germany
}

\date{\today}

\abstract{A recent paper demonstrated the existence of a secular aberration drift term in stellar proper motions that
    arises when transforming an astrometric catalogue defined for an observer at rest with respect to the solar system
    barycentre to some other reference frame in which, for example, the observer is at rest with respect to the Galactic
    centre. Such a transformation requires an accurate and precise estimate of the velocity of the solar system
barycentre. It was argued that the \gaia catalogue construction should account for this effect and also for the
aberrational effect due to acceleration of the solar system barycentre.}
{We argue that these two effects should not be accounted for in the construction of the \gaia astrometric catalogue.}
{We briefly review the \gaia catalogue reference frame, the concepts of stellar aberration and secular aberration
drift, and their observable consequences.}
{The \gaia catalogue is (and should be) constructed in the Barycentric Celestial Reference System: the reference system
    with the origin at the solar system barycentre as defined by the underlying solar system ephemerides. We explain
    that the \gaia catalogue is consistent with the International Celestial Reference System despite the presence of
    proper motion terms due to the acceleration of the solar system barycentre. We also explain why transformation of
the astrometry to a frame in which the observer is at rest with respect to the Galactic centre or distant universe is
not needed for the interpretation of stellar kinematics, and that there are practical concerns with such a
transformation.}
{The estimation of the velocity and acceleration of the solar system barycentre, although important as a matter of
scientific investigation, are not needed for the construction of the \gaia astrometric catalogue.}


\maketitle

\section{Introduction}
\label{sec:intro}

As a demonstration of the high quality of the astrometry published in \gaia Early Data Release 3 \citep{edr3paper,
missionpaper} an accompanying paper \citep{ssbaccel} presented a measurement of the acceleration of the solar system
barycentre (SSB) with respect to the rest frame of the distant universe, represented by a large number of quasars on the
whole sky. The result rests on the fact that the changing space velocity vector $\vssbc$ of the SSB relative to the rest
frame of the distant universe (i.e.\ the SSB acceleration ${\dif\over\dif t}\vssbc=\assbc$) leads to a systematic proper motion
pattern in quasar-like sources, which is measurable with \gaia. This proper motion pattern reflects the change in the
effect of aberration on source positions (due to the changing SSB velocity vector $\vssbc$) which is sometimes referred to
in the literature as `secular aberration drift'.

Recently, \citet{secabdrift} derived the presence of an additional secular aberration drift term caused by the changing
effect over time of stellar aberration for nearby sources (i.e.\ not at cosmological distances), in turn caused by the
changing line of sight direction to the sources (due to the motion of the sources with respect to the SSB).
\cite{secabdrift} consider an observer at the SSB at rest with respect to the extragalactic background but in the
development of their work adopt the results from \cite{reid2019} for the velocity of the SSB, which is actually the
velocity $\vssbg$ with respect to our Galaxy. According to their Eq.\ (16) the additional secular aberration drift term
causes a small change in the proper motion of a source which is proportional to $\vssbg$ and the proper motion itself. They
conclude that `Such a bias \ldots\ should therefore be considered in the reduction of \gaia astrometric data \ldots'.
Specifically they propose that stellar proper motions should be corrected for the secular aberration drift effects by:
1) removing the term caused by the SSB acceleration; and then 2) removing the additional secular aberration drift term
by iteratively determining from the astrometric catalogue the velocity $\vssbg$ of the SSB and the corrections to the
catalogue proper motions. The aim would be to make the (\gaia) stellar reference frame consistent with the
`extragalactic reference frame' (by which \citet{secabdrift} mean the rest frame of our Galaxy), and to improve our
understanding of stellar kinematics by the removal of apparent proper motion components.

We note that two velocities \vssbg and \vssbc differ both in their directions and absolute values. Velocity \vssbc is
the velocity of the SSB relative to the distant universe. In the standard cosmological model \vssbc can be measured e.g.
from the dipole of the Cosmic Microwave Background (CMB) radiation and was recently estimated as
$\left|\vssbc\right|\approx 370\kms$ \citep{2020A&A...641A...3P}. On the other hand, \vssbg is the velocity of the SSB
with respect to the rest frame of the Galaxy. It can be measured from the kinematics of the stars of our Galaxy and/or
from the accurate astrometric measurements of the radio source Sgr A* to give $\left|\vssbg\right|\approx 248\kms$
\citep{2020ApJ...892...39R}.

In this paper we explain why the corrections proposed by \cite{secabdrift} are, as a matter of principle, not pursued
in the construction of the \gaia catalogue, including for the upcoming data releases 4 and 5. Although the discussion in
the following sections is framed in terms of \gaia astrometry, we stress that this paper and its conclusions apply to
any modern astrometric data defined within the International Celestial Reference System.

\section{The \gaia catalogue reference frame}
\label{sec:gaiarefframe}

The astrometric data in the \gaia data releases are referred to the Barycentric Celestial Reference System
\citep[BCRS,][]{bcrsdef}. The origin of the BCRS coincides with the SSB as defined by the underlying solar system
ephemerides. The spatial axes of the BCRS are aligned with the International Celestial Reference System
\citep[ICRS,][]{icrsdef,icrsmeaning}. This choice of reference system means that \gaia source positions and proper
motions are those seen by an observer at rest at the origin of the BCRS and co-moving with that origin. We discuss the
consistency of the \gaia catalogue with the ICRS in \secref{sec:correctornot}.

\section{Aberration, secular aberration drift, and observable consequences}
\label{sec:abberation}

In this section we provide an alternative derivation of the \cite{secabdrift} result and clarify under what
circumstances the effects of aberration on source positions and proper motions can be observed.

The definition of stellar aberration in the Explanatory Supplement to the Astronomical Almanac \citep{explsupp} is as
follows: `The relativistic apparent angular displacement of the observed position of a celestial object from its
geometric position, caused by the motion of the observer in the reference system in which the trajectories of the
observed object and the observer are described.'\footnote{In the relativistic context one should also compute and remove
the effects of gravitational deflection of light before one gets the geometric positions of sources as seen by the
observer \citep{2003AJ....125.1580K}. However, in this discussion the effects of gravitation will be ignored.}.

This definition grasps the main point of aberration: the aberrational correction should be applied in order to work
directly with the (time-dependent) geometric positions of observed sources relative to a given observer as defined in
the selected reference system. The geometric positions both in the Newtonian and relativistic contexts are defined in
some selected reference system \sysA with spatial coordinates $\vect{x}=x^i$, $i=1,2,3$. In those coordinates the
trajectory of a source is $\vect{x}_{\rm src}(t)$ while that of the observer is $\vect{x}_{\rm obs}(t)$. Here $t$ is the
coordinate time of the selected reference system (reducing to the Newtonian absolute time in the Newtonian
approximation). Then the geometric position of the source is the unit vector $\vect{u}(t_{\rm obs})=\langle\vect{x}_{\rm
src}(t_{\rm em})-\vect{x}_{\rm obs}(t_{\rm obs})\rangle$, where $t_{\rm obs}$ is the time of observation, $t_{\rm em}$
is the time of emission of the signal that reaches the observer at time $t_{\rm obs}$, and $\langle\dots\rangle$ denotes
the normalization to a unit vector.

The standard parameterization of $\vect{u}(t_{\rm obs})$ gives the standard astrometric parameters of a source:
positions, proper motions, and parallaxes at a certain epoch. Detailed discussions of various aspects of the
parameterization can be found in \citet{2012A&A...538A..78L}, \citet{2014A&A...570A..62B}, and
\citet{2003AJ....125.1580K}. Along with the trajectories of the observer and the source, the astrometric parameters are
also defined in the reference system \sysA with time $t$ and spatial coordinates $\vect{x}$. The standard
parameterization of $\vect{u}(t_{\rm obs})$, which is intended for single stars, assumes that the spatial trajectory of
$\vect{x}_{\rm src}(t)$ is a linear function of time. This approximation should be sufficiently correct at least during
the time interval covered by astrometric data, but is usually considered true during a reasonable interval of time of a
few centuries discussed by modern astrometry. The requirement that $\vect{x}_{\rm src}(t)$ should be sufficiently close
to a linear function of time gives certain constraints that determine which reference systems can be used in astrometry
at a given level of accuracy and which cannot. In particular, to avoid rotation of the spatial coordinates the suitable
reference systems should be pseudo-inertial in the Newtonian approximation or its relativistic generalisations. In
addition the motion of the origin of the reference system should be suitable. For example, a reference system with the
origin at the centre of mass of the Earth with its quasi-periodic motion in the solar system with a period of about 1
year is not adequate for astrometry as soon as the observational accuracy exceeds several arcseconds. On the other hand,
a reference system with the origin at the SSB, as defined by a modern solar system ephemeris, can be used in astrometry
at the microarcsecond level: the non-linearity of stellar trajectories relative to the SSB due to the Galactic
gravitational forces is beyond reach even for \gaia astrometry \citep[][section 1.10, p.~102]{CTSR}. From hereon, the
origin of \sysA will be fixed at the SSB and its spatial coordinates will be denoted as $\vect{x}_\sysA$. This is the
reference system, in which the \gaia catalogue is constructed.

\subsection{Aberration and secular aberration drift}
\label{sec:abbanddrift}

Now, following \citet{secabdrift}, we consider another reference system \sysB with spatial coordinates
$\vect{x}_\sysB$\footnote{Specifically, \cite{secabdrift} consider a reference system \sysB in which the observer is at
    rest relative to the Galactic centre and thus moving with velocity $-\vssbg$ relative to \sysA, and they frame their
    discussion in terms of the velocity of \sysA with respect to \sysB. We choose to consider the velocity of \sysB
relative to \sysA in order to avoid the mental picture of the reference system \sysB somehow being more `fundamental'
than \sysA. According to the relativity principle the two points of view are equivalent.}. At some moment of time, the
origin of \sysB moves with respect to that of \sysA with a velocity \vb that is not necessarily a constant.
Reciprocally, the velocity of the origin of \sysA relative to \sysB at the same moment of time is $-\vb$. We assume here
that reference system \sysB can also be used in astrometry at the same level of accuracy as \sysA. In particular, this
is true if \vb is sufficiently close to a constant. The spatial coordinates of the reference systems \sysA and \sysB are
related to each other by a generalized Lorentz transformation \citep{bcrsdef}. For our discussion the Newtonian limit is
sufficient:
\begin{equation} \label{eq:xAtoxB} \vect{x}_\sysB=\vect{x}_\sysA-\vect{r}_{\sysA\sysB}(t)+{\cal O}(c^{-2})\,,
\end{equation}
where $\vect{r}_{\sysA\sysB}(t)$ is the position of the origin of \sysB in \sysA at the moment $t$, so that ${\dif\over
\dif t}\vect{r}_{\sysA\sysB}(t)=\vb$. \equrefalt{eq:xAtoxB} can be used to relate the trajectories of the sources and
the observer in the reference systems $\sysA$ and $\sysB$. The transformation of the trajectories allows one also to
transform the standard astrometric parameters of the sources. This means that the astrometric parameters obtained in
\sysA can be transformed to those measured in \sysB and vice versa without loss of information, provided \vb is known
accurately and to a precision that is compatible with the astrometric measurement precision. The latter point is
important for the discussion in \secref{sec:correctornot}.

On the other hand, the transformation~(\ref{eq:xAtoxB}) also gives the expression for the aberration of light when one
considers the light ray trajectory passing through the observer at the moment of observation from the apparent direction
$\vect{u}(t_{\rm obs})$: $\vect{x}_{\rm light}(t)=\vect{x}_{\rm obs}(t_{\rm obs})-c\,\vect{u}(t_{\rm obs})\,(t-t_{\rm
obs})+{\cal O}(c^{-1})$. Applying \eqref{eq:xAtoxB} to the light ray directions (and placing the observers in \sysA and
\sysB at the same physical location in space) allows one to relate the directions \ub and \ua, and using the
normalisation $\inprod{\vect{u}}{\vect{u}}=1$ one gets\footnote{In the Newtonian limit the aberrational formula is
    linear in the velocity, so that the aberrational correction due to the velocity of observer relative to the origin
of \sysA and due to \vb are additive. In the general relativistic case, \equref{eq:du2} should be replaced by a
relativistic formula which is non-linear with respect to the velocity and a more careful discussion of the velocity of
observer is needed. Here we assume that the accuracy of \equref{eq:du2} is sufficient for our discussion.}:
\begin{equation}
  \label{eq:du2}
  \du = \ub-\ua={1\over c}\vb-{1\over c}\inprodp{\ua}{\vb}\ua+{\cal O}(c^{-2})\,.
\end{equation}
\begin{figure}[t]
    \centering\includegraphics[width=0.8\linewidth]{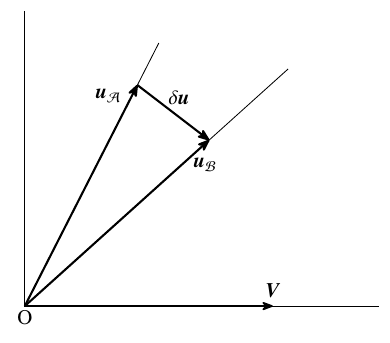}
    \caption{Schematic illustration of stellar aberration. The observer at O is moving with velocity \vb with respect to
    reference system \sysA and observes the object in the direction \ub. The geometric direction of the object in
    reference system \sysA is \ua. After Figure 7.3 in the Explanatory Supplement to the Astronomical Almanac
\citep{explsupp}.}
    \label{fig:schematic}
\end{figure}
The situation is shown schematically in~\figref{fig:schematic}, in which \ua and \ub are the unit vectors
representing the geometric directions for the same observation in reference systems \sysA and \sysB, respectively. In
particular, as \sysB we can choose the reference system moving relative to \sysA with velocities $-\vssbc$ or $-\vssbg$
discussed in \secref{sec:intro}. Thus, \sysB can be either at rest relative to the Galactic centre or relative to
the distant universe. Below we will sometimes speak of observers \sysA and \sysB whose positions coincide with the
origins of the respective reference systems at any moment of time.

The aberration effect in \equref{eq:du2} can change over time:
\begin{equation}
    \frac{\dif\du}{\dif t} = \frac{1}{c}\frac{\dif\vb}{\dif t} - \frac{1}{c}\inprodp{\ua}{\frac{\dif \vb}{\dif t}}\ua =
    \frac{\vect{a}}{c} - \frac{\inprod{\ua}{\vect{a}}}{c}\ua\,,
    \label{eq:abdrift}
\end{equation}
with $\dif\vb/\dif t = \vect{a}$ and where \ua is assumed not to change. This is referred to as `secular aberration
drift' and is what \cite{ssbaccel} used to estimate \vect{a} for the SSB. What \cite{secabdrift} pointed out is that the
source direction \ua also changes over time so that the more complete equation for the change in \du reads
\begin{equation}
    \frac{\dif\du}{\dif t} = \frac{\vect{a}}{c} - \frac{\inprod{\ua}{\vect{a}}}{c}\ua - 
    \frac{1}{c}\left[ \inprodp{\frac{\dif\ua}{\dif t}}{\vb}\ua + \inprodp{\ua}{\vb}\frac{\dif\ua}{\dif t}\right]\,.
\end{equation}
This shows that line of sight changes to sources (due to the change in geometric direction as observed in
reference system \sysA) can cause an additional secular aberration drift term. We now write the change of \ua as a
function of time in terms of the normal triad $[\vect{p}_0,\vect{q}_0,\vect{r}_0]$ \citep[cf.][section
3.2]{2012A&A...538A..78L} at the reference epoch $t=t_\mathrm{ep}$ (where $\ua(t_\mathrm{ep})=\vect{r}_0$) and the
proper motion $\vect{\mu}_0$ at $t_\mathrm{ep}$:
\begin{equation*}
    \left. \frac{\dif\ua}{\dif t}\right|_{t=t_\mathrm{ep}} = \vect{\mu}_0 =
        \vect{p}_0\,\mu_{\alpha*,0}+\vect{q}_0\,\mu_{\delta,0}\,,
\end{equation*}
with $\mu_{\alpha*,0}$ and $\mu_{\delta,0}$ the proper motions in right ascension and declination at $t_\mathrm{ep}$.
Here we assume that the observer is situated at the origin of \sysA and omit the parallax effect (which is not of
interest here) and the effect of the radial proper motion $\mu_r$ which does not lead to a change in \ua
\citep[cf.][their Eq.\ 4]{2012A&A...538A..78L}. Working out the additional term leads to the total secular aberration
drift
\begin{equation}
    \frac{\dif\du}{\dif t} = \frac{\vect{a}}{c} - \frac{\inprod{\vect{r}_0}{\vect{a}}}{c}\vect{r}_0 -
    \frac{1}{c}\left[ \inprodp{\vect{\mu}_0}{\vb}\vect{r}_0 + \inprodp{\vect{r}_0}{\vb}\vect{\mu}_0 \right]\,.
\end{equation}
Using the fact that the term $\inprodp{\vect{\mu}_0}{\vb}\vect{r}_0$ does not lead to a transverse proper motion, the
result is that the additional term in the secular aberration drift for stellar proper motions is:
\begin{equation}
    \Delta\vect{\mu} = -\frac{1}{c}\inprodp{\vect{r_0}}{\vb}\vect{\mu}_0\,.
    \label{eq:mudrift}
\end{equation}
This is the \cite{secabdrift} result which they derived by considering the effect of the epoch propagation of source
positions on the aberration in \equref{eq:du2} and then making an approximation for small epoch differences. Hence, the
full secular aberration drift in proper motion is:
\begin{equation}
    \Delta\vect{\mu}_\mathrm{sec} = \frac{\inprodp{\vect{a}}{\vect{p}_0}}{c}\vect{p}_0 +
    \frac{\inprodp{\vect{a}}{\vect{q}_0}}{c}\vect{q}_0 - \frac{1}{c}\inprodp{\vect{r_0}}{\vb}\vect{\mu}_0\,.
    \label{eq:fullmudrift}
\end{equation}
This equation shows the changes in the apparent proper motion components between the reference systems of observers
\sysA and \sysB due to the changes in \vb and \ua. We stress that the secular aberration drift terms simply result from
the transformation between reference systems \sysA and \sysB and thus strictly describe the differences between the
proper motions of the same sources observed in \sysA or \sysB. The differences are specific to the relative motion of
observers \sysA and \sysB described by \vb and \vect{a}.

\subsection{When are aberration effects observable in practice?}
\label{sec:observable}

We now ask whether aberration effects can be observed from reference system \sysA. This question arises because
\cite{secabdrift} consider the case where the astrometric source parameters as established in \sysA are to be
transformed to a reference system \sysB in which the secular aberration drift terms are removed, allowing for a better
study of, e.g., galactic dynamics. This requires that the aberration effects are established by the observer in \sysA
based on observations made in that same reference system \sysA. We first stress an important point again: the source
positions reported by an observer at rest in reference system \sysA are geometric in that reference system. Stellar
aberration effects that were not removed in the astrometric data processing, can only be observed if a certain pattern
in the geometric source positions \ua in \sysA can be related unambiguously to the effects of aberration.

We first consider the static aberration effect described by \equref{eq:du2} assuming \vb and \ua are unchanging. In this
case the observer \sysB is moving at constant velocity \vb and the stellar aberration leads to a static change in source
directions \ub compared to the directions \ua. The source positions as seen by observer \sysB will be distorted with
respect to the positions in \sysA, such that sources tend to concentrate toward the direction of motion \vb (see
\figref{fig:schematic}). Conversely, the source positions as seen by observer \sysA will be distorted with respect to
the positions in \sysB such that the sources tend to concentrate toward the direction $-\vb$. However this distortion
can only be measured by observer \sysA if they know the source directions in \sysB (which is an academic case only), or
if the source directions can be assumed to be uniform over the sky in a specific reference system. For distant QSOs the
assumption of a uniform sky distribution in the rest frame of the distant universe is reasonable \citep[and is used in
the literature to constrain the cosmic velocity \vssbc of the solar system, see e.g.][and references to that
work]{ellisbaldwin}. However, for stars this is far from true due to their highly non-uniform and imperfectly known
spatial distribution in the Milky Way. Hence, the static aberration effect is not observable in practice and is normally
ignored for stars.

A well-known observable aberration effect is that of the annual stellar aberration. Here the observer is orbiting the
SSB such that their velocity constantly changes direction, which causes a changing aberration effect for all stars. The
resulting apparent annual motion on the sky, with amplitude $\sim20$~arcsec, is the same for all stars (not dependent on
their distance). The most simple explanation for this pattern is the effect of aberration which can be calculated
precisely by transforming the directions seen by the observer in orbit around the SSB to the directions seen by an
observer at rest at the SSB. Indeed, in the \gaia astrometric data processing source positions are referred to the SSB
and corrections for the annual stellar aberration are applied. In this case the motion of \gaia with respect to the SSB
is very accurately known so the annual aberration effects can be taken out to sufficient accuracy.

Now we turn to \equref{eq:fullmudrift} and ask, are there geometric proper motion patterns observable in \sysA which can
most easily be explained as due to aberration? QSO-like sources are sufficiently distant that we can safely assume that
their proper motions are zero which means that for these sources the acceleration-related terms in
\equref{eq:fullmudrift} lead to a global proper motion pattern on the sky which can be exploited to estimate \assbc, as
done in \cite{ssbaccel}. The observability of the acceleration-related secular aberration drift thus relies on our
access to a sample of very distant sources which have zero geometric proper motions in the rest frame of the distant
universe (or any other reference system moving relative to that rest frame with a constant velocity). It is discussed in
great detail in \citep{ssbaccel} why no reasonable estimate of \assbc can be achieved using stars in our Galaxy: the
proper motions of stars are not predictable to the relevant level of accuracy.

QSO-like sources are of no use in observing the additional secular aberration drift term from \equref{eq:mudrift}
because for these objects $\vect{\mu}_0$ is zero. What sample of sources might allow this drift term to be observed?
Here we have to rely on nearby sources with significant proper motions in order to get a non-zero $\Delta\vect{\mu}$.
Figure~1 in \cite{secabdrift} illustrates that the factor $\inprod{\vect{r}_0}{\vb}$ shows a global pattern on the sky.
Possibly this pattern could be revealed if a sample of sources would be available for which we can on physical grounds
assume that these exhibit an ordered proper motion field $\vect{\mu}_0(\alpha,\delta)$ in \sysA or any other
well-defined reference system, which would be modulated in a detectable way by $\inprod{\vect{r}_0}{\vb}$. Such a sample
does not exist, as stars in the Milky Way exhibit a complex proper motion field, varying in direction and amplitude,
which completely hides the effect sought after. Hence, the additional secular aberration drift term in stellar proper
motions is in practice not observable in \sysA.

Moreover, as mentioned above the additional secular aberration drift term describes the difference between the proper
motions as observed in \sysA and \sysB. Not applying this additional aberrational effect does not alter the residuals of
the astrometric solution. One can always recompute one set of proper motions into another as soon as both reference
systems are strictly defined.

\section{Should the \gaia catalogue be corrected for secular aberration drift?}
\label{sec:correctornot}

Despite the effect described in \citet{secabdrift} not being observable one could still argue that we know that the SSB
moves and that with an estimate of \vb and \assbc one could transform the stellar proper motions to a reference system
in which the observer is at rest with respect to the Galactic centre or distant universe. Indeed \citet{secabdrift}
propose that the \gaia astrometry be corrected for the effects of secular aberration drift on the proper motions in
order to: 1) `ensure that the stellar reference frame aligns with the extragalactic reference frame'; and 2) `for
advancing our understanding of stellar kinematics'.

\subsection{The reference frame and secular aberration drift}
\label{sec:frame-and-ssbaccel}

Concerning the first point above, the extragalactic reference frame is not well defined in \citet{secabdrift}. For the
truly extragalactic reference frame, the rest frame of the distant universe and thus $\vb=-\vssbc$ should be used.
Instead, \citet{secabdrift} consider $\vb=-\vssbg$ and the rest frame of our Galaxy which is better referred to as the
`galactic reference frame'. Moreover, we note that the reference system suggested by \citet{secabdrift} is not defined
in a self-consistent way: the authors suggest to correct for the effects of \vssbg in proper motions, but ignore those
effects in positions; thus the standard parametrization of $\ub(t_{\rm obs})$ no longer works. In addition the
corrections for the effects of acceleration \assbc and \vssbg are inconsistent, as \assbc corresponds to \vssbc and not
\vssbg.

We emphasize that the \gaia catalogue (and thus the stellar reference frame) is constructed to be consistent with the
ICRS by aligning the \gaia catalogue positions with the International Celestial Reference Frame (ICRF), in the case of
\gdr{3} using ICRF3 sources present in the \gaia catalogue. In this process, the coordinates of the ICRF3 sources are
propagated to the epoch of the \gaia catalogue as prescribed by the ICRF3 \citep{icrf3}. To ensure that the \gaia-CRF is
globally non-rotating with respect to the distant universe, the rotational state of the \gaia catalogue is chosen such
that a large number of QSOs (not only the ICRF3 sources) show no common rotation in their proper motions
\citep[see][]{gcrf3}. We note here that the additional proper motions due to the acceleration of the SSB are
mathematically orthogonal to the frame rotation and do not influence the latter \citep[see][]{vsh2012}. The proper
motions of stars are not at all used to define the reference frame for \gaia.

The ICRS definition \citep{icrsdef} specifies the orientation of the axes to be fixed. Those axes can be materialized
both by celestial sources having constant positions as well as by sources having time-dependent positions. The only
necessary condition is that the positions of these sources are known at any moment of time in the ICRS axes. Naturally,
sources with no proper motion are more convenient for the materialization and until 2018 the ICRF realizations assumed
the positions of quasars as constant. With ICRF3 \citep{icrf3} the accuracy of VLBI astrometry has reached the level
where systematic proper motions due to the SSB acceleration should be taken into account \citep[see][section 3.2 and
references therein for a discussion and brief history of this topic]{icrf3}. Hence the choice made in the construction
of the ICRF3 to include a defining acceleration of the SSB in the data analysis (with an amplitude inferred from the
VLBI data) in order to account for the systematic proper motions. This allows to derive ICRF3 source positions
consistent with the ICRS axes and allows to correctly propagate the ICRF3 source positions to any desired epoch
\citep{icrf3}. Unlike the VLBI astrometry, in the \gaia data processing quasars are treated as normal astrometric
sources and their measured proper motions are published. Those proper motions contain both the systematic proper motions
due to the SSB acceleration and the individual apparent proper motions of these sources. The published proper motions
allow one to compute the positions of the \gaia quasars for any moment of time in the ICRS axes so that the orientation
of the \gaia reference frame is fully compatible with the ICRS. On the other hand, the proper motions published by \gaia
can be used by the community to measure the SSB acceleration, potentially improving the results.

What about the reference frame and the additional secular aberration drift term in stellar proper motions? As stressed
above, the positions and proper motions as listed in the \gaia catalogue reflect the geometric directions and their
change over time as seen by observer \sysA and thus form an accurate observational description of the celestial sphere.
An important role of the \gaia catalogue is in predicting future and past positions of sources on the sky in reference
system \sysA (epoch propagation) for applications such as stellar occultations, astrometric data reduction of surveys
anchored to \gaia, catalogue matching, and telescope pointing. If the \gaia catalogue were to list astrometric
parameters as seen by observer \sysB, i.e.\ removing the secular aberration drift terms, the use of this astrometry by
observer \sysA would actually lead to erroneous predictions of future or past positions of sources, as observed in
reference system \sysA, and thus spoil the above mentioned applications (unless one introduces a more complex algorithm
that first undoes the transformation from \sysA to \sysB, but this would be very error-prone in practice).

\subsection{Secular aberration drift and stellar kinematics}
\label{sec:stellar-kin}

The second point above raised by \cite{secabdrift} concerns the desire to study the kinematics of stars and the dynamics
of the Milky Way using proper motions free from aberration effects, i.e.\ the `true' proper motions as seen by the
observer \sysB with $\vb=-\vssbg$. However according to the relativity principle observer \sysB is no more fundamental
than observer \sysA, meaning that even if different observers moving at constant velocity relative to each other report
different stellar kinematics, the dynamics the observers infer should be the same. This is easily seen in the Newtonian
limit where the transformation between frames \sysA and \sysB amounts to a velocity offset only, which clearly will not
affect the dynamics of any stellar system under study (and we stress that the dynamics will also be the same if Lorentz
transformations are involved). Hence corrections for the term in \eqref{eq:mudrift} are not needed. The term due to the
SSB acceleration is sufficiently small at the current precision levels that it can be ignored. Also, its size is far
below many Galactic kinematic structures, as detailed in \cite{ssbaccel}, and can thus be ignored entirely.

We now discuss a more practical concern. The correction procedure proposed by \cite{secabdrift} depends on an accurate
knowledge of the SSB velocity (and acceleration). The velocity has to be estimated from the \gaia data to sufficient
accuracy to preserve the accuracy of the measured proper motions. The best relative proper motion uncertainties in the
\gdr{3} catalogue are of order $2.7\times10^{-6}$ and are expected to improve by a factor of $\sim8$ in the \gdr{5}.
This implies a knowledge of \vssbg at a level considerably better than 100\,m~s$^{-1}$, while modern estimates (based
also on \gaia data) are precise to the few km~s$^{-1}$ level, and are plagued by systematic uncertainties due to stellar
kinematics that deviate from a simple rotation curve \citep[see e.g.,][]{almannaei}. Moreover, if proper motion
corrections at the level of $\vssbg/c\sim8\times10^{-4}$ are considered important, then presumably corrections at the
same level to the other phase space coordinates (stellar positions) are also important. Correcting the celestial
positions for aberration subject to preserving the \gaia catalogue relative position uncertainties (of order
$5\times10^{-9}$ for faint stars to $5\times10^{-12}$ for bright stars) would require a knowledge of \vb down to the
mm~s$^{-1}$ level\footnote{Indeed the annual aberration corrections done as part of the \gaia astrometric data
processing rely on knowing the BCRS velocity of \gaia to about $1.0$~mm~s$^{-1}$ per component.}.

A second more fundamental practical concern is the following. Constructing the astrometric catalogue as seen in \sysB by
an observer at the SSB at rest with respect to our Galaxy (and thus moving with $\vb=-\vssbg$ with respect to \sysA)
necessitates introducing a priori our uncertain knowledge on the Milky Way galaxy, through \vssbg, \assbc, and the
assumption that $\assbc={\dif\over\dif t}\vssbc\approx{\dif\over\dif t}\vssbg$, into an observational catalogue which we
then want to use to study that same galaxy. This approach would spoil the purely observational result represented by the
\gaia catalogue, which is constructed within the very precisely defined BCRS and ICRS and lists well-defined astrometric
parameters derived from the measurements using the simplest assumption that stars follow straight line trajectories at
constant speed. We advocate that the estimation of \vssbc and \vssbg, as well as their derivatives, and possible
transformations of the astrometry to a different reference frame, belong to the realm of scientific investigation and
that therefore the \gaia catalogue should always be presented as a well-defined and accurate observational result, free
from assumptions about the Milky Way galaxy.

\section{Conclusion}
\label{sec:conclusion}

We have shown that the work presented in \cite{secabdrift}, even though the results are correct in the context of
considering what an observer at the SSB would see if at rest with respect to the Galactic centre, is of no practical
consequence for the construction and publication of the \gaia catalogue. The estimation of the velocity and acceleration
of the SSB, and possible transformations of the astrometry to a frame in which the observer is at rest with respect to
the Galactic centre or distant universe, are subjects of scientific investigation, not \gaia astrometric catalogue
construction.

\begin{acknowledgements}
    The authors thank Jos de Bruijne for triggering the discussion that resulted in this paper, and the referee for a
    critical report that prompted a thorough rethinking and clarification of the issues to be addressed. 
    \\
    This work has made use of data from the European Space Agency (ESA) mission {\it Gaia}
    (\url{https://www.cosmos.esa.int/gaia}), processed by the {\it Gaia} Data Processing and Analysis Consortium (DPAC,
    \url{https://www.cosmos.esa.int/web/gaia/dpac/consortium}). Funding for the DPAC has been provided by national
    institutions, in particular the institutions participating in the {\it Gaia} Multilateral Agreement.
    \\
    While preparing this work we made use of the following software: Astropy, a community-developed core Python
    package for Astronomy \citep[\url{https://www.astropy.org}]{astropy:2013, astropy:2018, astropy:2022}, IPython
    \citep[\url{https://ipython.org/}]{ipython:2007}, Jupyter (\url{https://jupyter.org/}), Matplotlib
    \citep[\url{https://matplotlib.org}]{matplotlib:2007}, NumPy \citep[\url{https://numpy.org}]{numpy:2020}, TOPCAT
    \citep[\url{https://www.starlink.ac.uk/topcat/}]{topcat:2005}, and the Ti\emph{k}Z and PGF packages \cite{tantau2024}.
    \\
    This research has made use of the Astrophysics Data System, funded by NASA under Cooperative Agreement
    80NSSC21M00561, and the SIMBAD database \citep[SIMBAD;][]{cds2000} operated at the CDS, Strasbourg, France
    (\href{https://cds.u-strasbg.fr/}{CDS}).
\end{acknowledgements}

\bibliographystyle{aa}
\bibliography{refs-sad}

\end{document}